\documentstyle[aps,prd]{revtex}

\begin{document}
\tighten

\title{Nonlinear bulk viscosity and inflation}

\author{Roy Maartens\footnote{maartens@sms.port.ac.uk}}

\address{Department of Mathematics, University of
Natal, Durban 4041, South Africa \\
and School of Mathematical Studies, Portsmouth University,
Portsmouth PO1 2EG, Britain\footnote{permanent address}}

\author{Vicen\c c M\'endez\footnote{vicenc@ulises.uab.es}}

\address{Departament de Fisica, Universitat Autonoma de Barcelona,
E--08193 Bellaterra, Spain}

\date{November 1996}

\maketitle

\begin{abstract}

We develop a nonlinear generalisation  
of the causal linear thermodynamics of bulk viscosity, incorporating
positivity of the entropy production rate and the effective specific 
entropy. The theory is applied to viscous fluid inflation (which is
necessarily far from equilibrium), and we find
thermodynamically consistent
inflationary solutions, both exponential and power--law.

\end{abstract}


\section{Introduction}

Scalar dissipative processes in cosmology may be
treated via the relativistic theory of bulk viscosity (see \cite{m1},
\cite{z1} and references cited there). The causal and stable
thermodynamics of Israel/ Stewart provides a satisfactory
replacement of the unstable and non--causal theories of Eckart
and Landau/ Lifshitz. However, it shares with
these theories the limitation that it is based on assuming
small departures from equilibrium,
so that the transport equation is linear in the bulk viscous pressure. 
While this linear assumption
is reasonable for many cosmological and astrophysical
situations, there may be
processes which involve large deviations from equilibrium. Such
deviations are likely to lead to a breakdown of the theory. (This
is known to occur for heat flux \cite{hl1}, \cite{cl}.)

For example, if viscosity--driven inflation occurs (leaving aside
for the moment various
questions about the hydrodynamical consistency of such models), then 
this necessarily involves nonlinear bulk viscous pressure \cite{m1}. 
Any
application of the linear theories to viscous inflation 
(see \cite{m1} for a review),
unavoidably requires an assumption, which is usually not made
explicit, that linear thermodynamics remains valid well beyond the
regime in which it is derived. The results of such a linear analysis
are questionable. The alternative is to develop
a nonlinear generalisation of the Israel/ Stewart theory
for a more satisfactory model of viscosity--driven inflation.
The goal of studying viscosity--driven inflation is to find 
inflationary solutions arising strictly from `ordinary' matter in a 
far--from--equilibrium state, without invoking scalar fields.
Of course, such models also need to address the question of how
the observed density perturbation spectrum can be produced if
one eliminates the usual scalr inflaton. This important
problem is not addressed here.

We present a
phenomenological model of nonlinear bulk viscosity and apply
it to viscous inflation. In the absence of 
a suitable microscopic foundation
for a nonlinear theory, we propose a model which at least satisfies
reasonable physical constraints, in that it:
\begin{enumerate}
\item
approaches the
Israel/ Stewart theory in the linear regime (i.e. small deviations
from equilibrium), thus ensuring causal
and stable behaviour in that regime;
\item
ensures that the entropy 
production is non--negative, i.e. the second law of thermodynamics
is satisfied;
\item
imposes an upper limit on the
bulk viscous stress, analogous to existing nonlinear generalisations 
of the heat flow equation \cite{jcl}. 
\end{enumerate}

In Section II we discuss briefly the problem of
expanding fluids that are far from equilibrium, especially the case
of inflationary expansion. 
On the basis of the above guiding principles, we then
put forward in Section II
the mathematically simplest form for a nonlinear transport
equation, and consider
some of its properties. Hopefully this simple model can illuminate
some of the overall features of a microscopically--based nonlinear
theory, which would probably depend heavily on the particular
interactions involved. 
We find in Section III exact solutions in a flat FRW universe,
for simple and consistent
thermodynamic coefficients and equations of state.
Provided the thermodynamic parameters satisfy certain conditions,
the model admits thermodynamically consistent inflationary 
solutions. Although viscous fluid inflation is the only application
considered here, the nonlinear theory could in principle be applied
to other far--from--equilibrium processes.

\section{Transport equation for bulk viscosity}

The hydrodynamic description which is 
implicit in fluid dynamics, requires that the 
mean interaction time $t_c$
of fluid particles should be much
less than any characteristic macroscopic
time--scale. In cosmology, this requires
\begin{equation}
t_c \ll H^{-1}\,,
\label{1}\end{equation}
where $H$ is the Hubble expansion rate. Condition (\ref{1}) means that
the fluid is able to adjust to the cooling caused by expansion, and
to establish a state
amenable to a hydro--thermodynamical description.
In particular, if the fluid is in or close to equilibrium, then
(\ref{1}) should ensure that the fluid has a well--defined
local temperature. 

For a fluid far from equilibrium, there will not in general be a
well-defined temperature. For example, very high and rapidly
varying temperature
gradients could give rise to large heat flux, without a meaningful
temperature at any event. This could happen without 
violating (\ref{1}).

If we focus on scalar dissipative effects,
in particular ruling out heat flux, then large deviations from
equilibrium arise from large bulk viscous stresses, 
i.e. $|\Pi|\gtrsim p$, where $p$ is
the local equilibrium pressure, and
the effective non--equilibrium pressure is
\begin{equation}
p_{e\!f\!f}=p+\Pi \,.
\label{2}\end{equation}
In an expanding fluid, the dissipation due to $\Pi$ leads to a 
decrease in kinetic energy and therefore in pressure, so that
$\Pi\leq0$.

Qualitatively,
bulk viscosity is the macroscopic expression of 
microscopic `frictional' effects that arise in mixtures \cite{ui},
\cite{z2}.
The bulk viscosity tends to be largest when the contrast between
components of the mixture 
(specifically, the contrast in cooling rates)
is greatest. For example:
\begin{enumerate}
\item
a mixture of
ultra--relativistic and non--relativistic particles in an 
ideal gas in the intermediate regime $mc^2\sim kT$;
\item
a mixture of
photons with long mean free path, and
thermalised non--relativistic electrons and protons,
whose mean free path is very short (radiative hydrodynamics);
\item
a mixture of gauge bosons which have acquired mass after a
phase transition in the very early universe, and massless and
effectively massless thermalised species.
\end{enumerate}

Bulk viscosity allows us to describe a mixture of different species 
effectively as a single fluid, provided 
a hydrodynamic description is reasonable. For example, in 
radiative hydrodynamics, if the thermalised matter dominates
the energy density, then since it obeys (\ref{1}), we expect that a
single--fluid model is reasonable, even though the photon component
strongly violates (\ref{1}). Far from equilibrium, the fluid model
itself may break down, in the sense that (\ref{1}) may 
not be satisfied
by any interactions. We will assume that the deviations from 
equilibrium are not such as to cause a breakdown in the fluid
description. The non--equilibrium conditions could lead to a
weakening of (\ref{1}), i.e. to $t_c<H^{-1}$. 

In the case of viscous fluid inflation (discussed in more detail
in the following section), it is difficult to see how short--range
interactions in the fluid could maintain a rate that is greater than
the inflationary expansion rate. Fluid particles are being separated
at tremendous speed, so that (\ref{1}) should be strongly violated
and a hydrodynamic description should break down.
Any consistent fluid model of inflation needs to provide a way around 
this problem. One possibility could be the existence of
{\em long--range} interactions, possibly involving a coupling of
particle interactions to gravity, whose rate remains higher than
the inflation rate. 
In our subsequent discussion of viscous fluid inflation, we 
will assume that some such mechanism exists to ensure consistency
of the model. (Without such a mechanism being explicit, the model
remains phenomenological.)
Our task is then to provide a nonlinear transport
equation to deal with the large bulk viscous stress involved in
inflation.

A covariant approach to the causal thermodynamics of
relativistic fluids \cite{is}, \cite{hl} 
is based on the hydrodynamic tensors
$n^\alpha$ (particle number 4--current), $T_{\alpha\beta}$ 
(energy--momentum tensor) and $S^\alpha$ (entropy 4--current), which
are subject respectively to number and energy--momentum conservation
\begin{equation}
n^\alpha{}_{;\alpha}=0\,,\quad T^{\alpha\beta}{}{}_{;\beta}=0\,,
\label{3}\end{equation}
and to non--negative entropy production rate (the second law of
thermodynamics)
\begin{equation}
S^\alpha{}_{;\alpha}\geq 0 \,.
\label{4}\end{equation}
The entropy 4--current is assumed to be algebraically determined by
$n^\alpha$ and $T_{\alpha\beta}$ alone:
\begin{equation}
S^\alpha=S^\alpha(n^\beta,T^{\mu\nu}) \,.
\label{5}\end{equation}
This is a reasonable assumption near equilibrium, and is 
well--motivated by kinetic theory \cite{is}. Far from equilibrium,
(\ref{5}) could be a drastic assumption, in the sense that the
full structure of non--equilibrium states is unlikely to be captured
by only the first two moments of the particle distribution. However,
for only scalar dissipation, it may be a less restrictive assumption.
Furthermore, in homogeneous FRW spacetimes, the assumption that
$S^\alpha$ does not depend on spatial gradients of the moments
will not be restrictive.
In any case, without a clear alternative, we can begin by assuming
that (\ref{5}) is also valid for nonlinear scalar deviations from
equilibrium.

At each event, one can define a local reference equilibrium state,
characterised by number density $n$, energy density $\rho$, pressure
$p$, specific entropy $S$ and 4--velocity $u^\alpha$. 
The arbitrariness in the reference state can be used to match $n$ and
$\rho$ to the actual state, in the particle frame, which is 
characterised by vanishing particle flux, so that
\begin{equation}
n^\alpha=n u^\alpha \,.
\label{6}\end{equation}
The energy--momentum tensor for scalar dissipation is then
\begin{equation}
T_{\alpha\beta}=\rho u_\alpha u_\beta+(p+\Pi)h_{\alpha\beta} \,,
\label{7}\end{equation}
where $h_{\alpha\beta}=g_{\alpha\beta}+u_\alpha u_\beta$ is the
projection tensor, with $g_{\alpha\beta}$ the metric.
In the absence of vector and tensor dissipation, the entropy
4--current will be of the form
\begin{equation}
S^\alpha=S_{e\!f\!f}\,n^\alpha \,,
\label{8}\end{equation}
where $S_{e\!f\!f}$ is the effective, non--equilibrium 
specific entropy.

The conservation equations (\ref{3}) for (\ref{6}) and
(\ref{7}) are
\begin{eqnarray}
\dot{n}+3Hn=0 \label{9} \,,\\
\dot{\rho}+3H(\rho+p+\Pi)&=&0 \,,\label{10}\\
(\rho+p+\Pi)\dot{u}_\alpha+h_\alpha{}^\beta(p+\Pi)_{,\beta}&=&0 \,,
\label{11}\end{eqnarray}
where $3H=u^\alpha{}_{;\alpha}$ is the average volume rate of 
expansion ($H=$ Hubble rate in FRW spacetime), 
and $\dot{u}_\alpha=u_{\alpha;\beta}u^\beta$ is the
4--acceleration.

The transport equation for $\Pi$ arises from imposing the second
law (\ref{4}) on (\ref{8}) -- which requires us to specify the form
of $S_{e\!f\!f}$. In the Israel/ Stewart theory
\begin{equation}
S_{e\!f\!f}=S-\left({\tau\over 2nT\zeta}\right)\Pi^2 \,,
\label{12}\end{equation}
where $\zeta(\rho,n)$ is the bulk viscosity and
$\tau(\rho,n)$ is the characteristic time for linear
relaxational or transient effects.
The local equilibrium variables $S$ and $T$ satisfy the Gibbs 
equation 
\begin{equation}
TdS=(\rho+p)d\left({1\over n}\right)+{1\over n}d\rho \,,
\label{13}\end{equation}
which implies, using (\ref{9}) and (\ref{10}), that
\begin{equation}
\dot{S}=-{3H\Pi\over nT} \,.
\label{13a}\end{equation}
It is a feature
of the Israel/ Stewart theory that dissipative contributions to the
entropy 4--current are quadratic. Any linear contribution is ruled out
by the requirement that the entropy density be locally a maximum in
equilibrium. 

In the Israel/ Stewart theory, the deviations from
equilibrium are small, i.e. 
\begin{equation}
|\Pi|\ll p \,.
\label{14}\end{equation}
Since $nT\sim p$ and
$\tau/\zeta\sim\rho^{-1}\sim p^{-1}$, it follows from (\ref{12}) 
and (\ref{14}) that $S-S_{e\!f\!f}\ll 1$, so that (\ref{14})
prevents the effective specific entropy from becoming negative.
We will not impose the constraint (\ref{14})
on $|\Pi|$. However,
for simplicity, we will {\em retain} the
quadratic form (\ref{12}), i.e. we will not add cubic and higher 
contributions, since there is no clear indication as to what such
contributions would be -- and since
they would introduce further, unknown
coupling coefficients. Since we require non--negative
$S_{e\!f\!f}$, a consequence of
adopting (\ref{12}) in the nonlinear case is that the bulk stress
is bounded:
\begin{equation}
S_{e\!f\!f}\geq0\quad\Leftrightarrow\quad
\left|\Pi\right|\leq\left|\Pi\right|_{m\!a\!x}=\sqrt{{2nTS\zeta\over
\tau}} \,.
\label{14a}\end{equation}
Apart from this new feature, the main
difference between the nonlinear
theory presented here, and the linear Israel/ Stewart theory, reduces
to {\em the different ways in which the second law is imposed}.

By (\ref{9}), (\ref{10}) and (\ref{13}), we find that (\ref{8})
and (\ref{12}) lead to
\begin{eqnarray}
S^\alpha{}_{;\alpha}&=&-{\Pi\over T}{\cal X}\quad\mbox{where}
\nonumber\\
{\cal X}&=& 3H+{\tau\over\zeta}\dot{\Pi}+
{\tau\over 2\zeta}\Pi
\left(3H+{\dot{\tau}\over\tau}-{\dot{\zeta}\over\zeta}
-{\dot{T}\over T}\right) \,.
\label{15}
\end{eqnarray}
In the Israel/ Stewart theory, the second law (\ref{4}) is satisfied
by assuming a {\em linear} relation between the thermodynamic `force'
${\cal X}$ and the thermodynamic `flux' $\Pi$, i.e.
\begin{equation}
\Pi=-\zeta{\cal X}\quad\mbox{(Israel/ Stewart)}\,.
\label{16}\end{equation}
(The linearity may be justified via near--equilibrium
kinetic theory \cite{is}.)
Formally, it follows
from (\ref{16}) that $|\Pi|$ is unbounded in response to growth in
${\cal X}$. 
The near--equilibrium condition (\ref{14}) has to be
taken as an additional, extraneous constraint on the transport 
equation (\ref{16}).
This is a sign of the limitations of the theory arising
from its linearity. Beyond near--equilibrium conditions, the linear
transport equation (\ref{16})
cannot be expected to give reasonable predictions.

The nonlinear generalisation of (\ref{16}) which we propose 
must be consistent with the upper bound (\ref{14a})
on $|\Pi|$, so that while $|\Pi|/p$ is
not restricted to be small, it does reach a limit as the 
thermodynamic force is increased without bound. This seems to be
physically reasonable, and is in line with existing phenomenological
generalisations of the heat transport equation \cite{jcl}. The simplest 
generalisation of (\ref{16}) that we could find
with this property, and the other
properties listed in the introduction, is
\begin{equation}
\Pi=-{\zeta{\cal X}\over 1+\tau_*{\cal X}}\quad\mbox{(nonlinear)} \,,
\label{17}\end{equation}
where $\tau_*\geq0$ is a characteristic time--scale for nonlinear
effects. Nonlinear effects are significant for 
${\cal X}\gtrsim \tau_*^{-1}$.

For small thermodynamic force ${\cal X}$, equivalently 
small $|\Pi|$, (\ref{17}) tends to the 
form (\ref{16}), and the linear Israel/ Stewart theory is recovered.
For $|{\cal X}|\rightarrow\infty$, 
$-\Pi$ tends to the upper bound $\zeta/\tau_*$.
It follows from (\ref{15}) and (\ref{17}) that 
\begin{equation}
S^\alpha{}_{;\alpha}= n\dot{S}_{e\!f\!f}
= {\Pi^2\over T\zeta}\left[1+{\tau_*\over\zeta}\Pi\right]^{-1}\,.
\label{19}\end{equation}
This means that the second law holds identically by virtue of the
upper bound on the bulk stress:
\begin{equation}
S^\alpha{}_{;\alpha}\geq0\quad\Leftrightarrow\quad
-\Pi\leq{\zeta\over\tau_*} \,.
\label{18}\end{equation}
As $-\Pi$
approaches $\zeta/\tau_*$, the entropy 
production rate (\ref{19}) tends to infinity, providing a barrier
against values $-\Pi>\zeta/\tau_*$. Note also that (\ref{19}) implies
that the rate of growth of effective specific entropy is greater in
the nonlinear than in the linear theory. The condition (\ref{14a})
implies by (\ref{18}) that
\begin{equation}
\tau_*\geq\sqrt{{\zeta\tau\over 2 nTS}} \,.
\label{21}\end{equation}

By (\ref{15}), the detailed form of the
transport equation (\ref{17}) is
\begin{eqnarray}
&&\tau\dot{\Pi}\left(1+{\tau_*\over\zeta}\Pi\right)+\Pi\left(1+
3\tau_*H\right) \nonumber\\
&&{}=-3\zeta H-{\textstyle{1\over2}}\tau\Pi\left[3H+{\dot{\tau}\over
\tau}-{\dot{\zeta}\over\zeta}-{\dot{T}\over T}\right]\left(1+
{\tau_*\over\zeta}\Pi\right) \,.
\label{20}\end{eqnarray}
The Israel/ Stewart 
theory, in its full (non--truncated) form \cite{m1}, 
\cite{z1}, 
is the limiting case $\tau_*=0$, when the terms
in round brackets in (\ref{20}) all reduce to 1.

The thermodynamic coefficients $\zeta$ and $\tau$ are known from
the linear theory, based on kinetic theory arguments \cite{is}. 
For the
nonlinear coefficient $\tau_*$ we do not have a corresponding
microscopic derivation. However the phenomenological theory presented
here imposes on $\tau_*$ the limit (\ref{21}), which
is determined by the linear
coefficients and local equilibrium variables.

\section{Thermodynamically consistent inflation}

We specialise now to a flat FRW universe
$$
ds^2=-dt^2+a(t)^2\left(dx^2+dy^2+dz^2\right) \,,
$$
where $H=\dot{a}/a$. The transport equation (\ref{20}) implies, with
the Einstein field equations, an evolution equation for $H$, once
we specify the thermodynamic coefficients and equations of state.
The field equations are
\begin{eqnarray}
H^2 &=& {\textstyle{1\over3}}\rho \,,\label{22}\\
\dot{H}+H^2 &=& -{\textstyle{1\over6}}\left(\rho+3p+3\Pi\right) \,.
\label{23}
\end{eqnarray}
For the equilibrium pressure we impose the linear barotropic equation 
of state
\begin{equation}
p=(\gamma-1)\rho \,,
\label{24}\end{equation}
where $\gamma$ is a constant, $1\leq\gamma\leq2$. 
If $T$ is also barotropic, $T=T(\rho)$,
then the integrability condition
$\partial^2S/\partial\rho\partial n=
\partial^2S/\partial n\partial\rho$ of the Gibbs equation (\ref{13})
leads to \cite{m2}
\begin{equation}
T\propto \rho^{(\gamma-1)/\gamma} \,.
\label{25}\end{equation}
We note that the ideal gas law $p=nT$ is {\em incompatible} with 
barotropic temperature and the linear
barotropic equation of state (\ref{24}), unless $\Pi=0$ \cite{mt}. 
This follows
from (\ref{25}) -- which is a consequence of (\ref{24}) 
and $T=T(\rho)$ -- and the
conservation equations (\ref{9}) and (\ref{10}). 

The linear relaxation time is related to the bulk viscosity by
\cite{m2}
\begin{equation}
\tau={\zeta\over v^2(\rho+p)}={\zeta\over v^2\gamma\rho} \,,
\label{26}\end{equation}
where $v$ is the dissipative contribution to the speed of sound
$V$, so that $V^2=c_s^2+v^2$, with $c_s$ the adiabatic contribution.
By causality, $V\leq1$, so that
\begin{equation}
v^2\leq 1-c_s^2=2-\gamma \,,
\label{26a}\end{equation}
where we used (\ref{24}). We assume that $v$ is constant, like
$c_s$ is when (\ref{24}) holds.
The bulk viscosity itself is often taken to
be of the simplified form $\zeta\propto\rho^{q/2}$
($q$ constant), which should not be unreasonable in 
FRW spacetime with (\ref{24}). By (\ref{22}) we have
\begin{equation}
\zeta=\alpha H^q \,,
\label{27}\end{equation}
where $\alpha$ ($\geq0$) is constant.
Finally, we need to specify the
nonlinear characteristic time $\tau_*$. The simplest possibility
appears to be
\begin{equation}
\tau_*=k^2\tau \,,
\label{28}\end{equation}
where $k$ is constant, and $k=0$ is the linear (Israel/ Stewart)
case. This ansatz is subject to the condition (\ref{21}), which does
not take a simple form in general.

From (\ref{22}) -- (\ref{24}) we get
\begin{equation}
\Pi=-2\dot{H}-3\gamma H^2 \,.
\label{29}\end{equation}
Together with (\ref{25}) -- (\ref{28}), this brings (\ref{20}) to
the form
\begin{eqnarray}
\left[1-{k^2\over v^2}-\left({2k^2\over3\gamma v^2}\right){\dot{H}
\over H^2}\right]\left\{\ddot{H}+3H\dot{H}+\left(
{1-2\gamma\over\gamma}\right){\dot{H}^2\over H}+{\textstyle{9\over4}}
\gamma H^3\right\} && \nonumber\\
{}+{3\gamma v^2\over2\alpha}\left[1+\left({\alpha k^2\over\gamma v^2}
\right)H^{q-1}\right]H^{2-q}\left(2\dot{H}+3\gamma H^2\right)
-{\textstyle{9\over2}}\gamma v^2H^3=0 \,. &&
\label{30}\end{eqnarray}
This is the fundamental dynamical equation for nonlinear bulk
viscosity in a flat universe. Once (\ref{30}) is solved for $H$,
$\rho$ is determined by (\ref{22}) and $\Pi$ by (\ref{29}).
When $k=0$, the terms in square brackets all reduce to 1, and we
recover the dynamical equation for the (non--truncated) 
Israel/ Stewart theory \cite{m1}.\footnote{Note that \cite{m1}
effectively assumes
$v^2=1/\gamma$, and takes the exponent in (\ref{25}) to be a
free parameter $r$; one must set $r=(\gamma-1)/\gamma$ in \cite{m1}.}

Viscous inflation has been discussed in a number of papers
(see \cite{m1} for a review), both as a phenomenological model of
quantum particle and string creation effects, and as a fluid
alternative to scalar--field inflation. 
Particle production is formally equivalent to an effective 
bulk viscosity \cite{z1}, 
but we will confine ourselves here to discussing fluid
models, in which the bulk viscous stress $|\Pi|$
has become large enough to
make the effective pressure (\ref{2})
negative and initiate inflation, and where net particle creation and
decay occur only during reheating,
at the end of inflation. (Causal linear thermodynamics, incorporating
net particle production, may be used as a phenomenological model of
reheating \cite{zpmc}.)
As discussed in the previous
section, we assume that long--range 
interactions or some other microscopic mechanism are present to 
ensure that a hydrodynamical description remains reasonable during
inflation.

The condition for inflation $\ddot{a}>0$ implies by (\ref{23}) that
\begin{equation}
-\Pi>p+{\textstyle{1\over3}}\rho \,.
\label{31}\end{equation}
It is clear from (\ref{31}) that viscous fluid inflation is not
close to equilibrium, i.e. it strongly
violates the condition (\ref{14}), and
we need to use the generalised nonlinear equation (\ref{30}).
We consider first whether (\ref{30}) admits exponential inflation, and
if so, whether the solutions conform to the thermodynamic conditions
discussed above.

Setting $H=H_0$, (\ref{30}) gives
\begin{equation}
H_0^{1-q}={\alpha\over 2\gamma v^4}\left(2v^2-1\right)\left(
v^2-k^2\right) \,.
\label{32}\end{equation}
By (\ref{29}) and (\ref{22}), $\Pi=-3\gamma H_0^2=-\gamma\rho$. Then
the condition (\ref{18}) implies, using (\ref{26}) and (\ref{27}), 
a limit on $k$:
\begin{equation}
k\leq v \,.
\label{33}\end{equation}
In order to check the condition (\ref{21}), we evaluate the specific 
entropy. Integrating (\ref{13a}) with $T=T_0=$ constant:
\begin{equation}
S=S_0+\left({3\gamma H_0^2\over n_0 T_0}\right)e^{3H_0t} \,,
\label{34}\end{equation}
where $n=n_0a^{-3}$.
Then (\ref{12}) leads to
\begin{equation}
S_{e\!f\!f}=S_0+\left({3\gamma H_0^2\over 2n_0T_0v^2}\right)
\left(2v^2-1\right)e^{3H_0t} \,.
\label{35}\end{equation}
Thus we require $v^2\geq {1\over2}$.
It follows from (\ref{32}), (\ref{33}) and (\ref{26a}) 
that exponential inflation 
is possible provided 
\begin{equation}
k<v\,,\quad {\textstyle{1\over2}}<v^2\leq 2-\gamma \,,
\label{36}\end{equation}
and then 
the entropy generation and effective specific entropy are positive.
Thus the solutions are thermodynamically consistent.  Note that
$\gamma<{3\over2}$. Exponential growth of $S$ is also
found in the linear theory \cite{m1} -- but the advantage here
is that the theory is nonlinear, and we consider the effective
specific entropy, which is more suitable far from equilibrium.
The apparent closeness of the linear and nonlinear results is an
artefact of the constancy of $H$ and thereby of $\rho, T, \Pi$.
As shown in \cite{m1}, it is possible to generate a large amount
of entropy via the exponential form (\ref{35}). 

Following the qualitative analysis techniques for the Israel/ Stewart
solutions \cite{mt}, \cite{cvm}, we find that the stability
properties of the solution (\ref{32}) depend on the parameter
\begin{equation}
\omega={\left(2v^4+v^2-2k^2\right)^2\over
4\gamma \left(v^2-k^2\right)^2\left(2v^2-1\right)}>0 \,.
\label{37}\end{equation}
Then the fixed point $(0,H_0)$ in the phase plane $\{(\dot{H},H)\}$ 
has the following properties:\\
for $1-\omega\leq q<1$, it is an asymptotically stable
node;\\ 
for $q>1$, it is a saddle point; \\
for $q<1-\omega$, it is an asymptotically stable focus.

Thus the solution is an attractor for $q<1$.

When $q=1$, exponential inflation can occur with $H_0$ arbitrary,
provided that the bulk viscous parameter satisfies
\begin{equation}
\alpha={2\gamma v^4\over\left(2v^2-1\right)\left(v^2-k^2\right)} \,,
\label{38}\end{equation}
as follows from (\ref{32}). Of more interest is the existence
when $q=1$
of power--law solutions $a=a_0 t^N$, which are inflationary for $N>1$. 
By (\ref{30}), these solutions exist if
\begin{eqnarray}
&& 27\gamma^3\left[\alpha\left(1-2v^2\right)\left(v^2-k^2\right)+2
\gamma v^4\right]N^3  \nonumber\\
&&{}+18\gamma^2\left[3\alpha k^2-2\gamma v^4-2\alpha v^2
\left(1+k^2\right)\right]N^2  \nonumber\\
&&{}+ 12\alpha\gamma\left(v^2-3k^2\right)N+8\alpha k^2=0 \,.
\label{39}\end{eqnarray}
Clearly there are solutions of this cubic for a wide range of
thermodynamic parameters. However, we are concerned to identify the
thermodynamically consistent solutions, for which the conditions are
stringent. 

By (\ref{29}), $\Pi=-N(3\gamma N-2)t^{-2}$, and together with
(\ref{22}), (\ref{25}), (\ref{26}) and (\ref{28}), this shows that
the condition (\ref{18}) for positive entropy generation is
\begin{equation}
k\leq\left({3\gamma N\over 3\gamma N-2}\right)^{1/2}v \,.
\label{40}\end{equation}
Integrating (\ref{13a}) we get 
$$
S=S_0+\left[{\gamma\left(3N^2\right)^{1/\gamma}\over n_0T_0a_0^3}
\right]t^{(3\gamma N-2)/\gamma} \,,
$$
and then (\ref{12}) gives
\begin{equation}
S_{e\!f\!f}=
S_0+\left[{\left(3N^2\right)^{(1-\gamma)/\gamma}
\over 6\gamma v^2n_0T_0a_0^3}\right]
\left\{9\gamma^2\left(2v^2-1\right)N^2+12\gamma N-4\right\}
t^{(3\gamma N-2)/\gamma} \,.
\label{41}\end{equation}
The effective specific entropy is positive if the
$N$--quadratic in braces is positive. This requires that
\begin{equation}
v\leq {1\over\sqrt 2} \,,
\label{42}\end{equation}
and it places an upper limit on $N$, so that
\begin{equation}
{2\over3\gamma}<N\leq {2\over3\gamma\left(1-\sqrt{2}\, v\right)} \,.
\label{43}\end{equation}
(The lower limit ensures that $\Pi<0$.)

Thus the constraints (\ref{40}), (\ref{42}) and (\ref{43}) must be
satisfied for any solution of (\ref{39}) to be thermodynamically
consistent. We present a simple example to show that consistent
inflationary solutions do exist:
$$
\gamma={4\over3}\,,\quad v^2={1\over2}=k^2\,,
\quad\alpha={67\over 4}\,,\quad N=2 \,.
$$
\[ \]

In conclusion, we have argued for a nonlinear generalisation
of causal linear thermodynamics in order to describe viscous
inflation (without particle production). A simple phenomenological
theory, supplemented by a simple
model for the thermodynamic coefficients and equations of state,
leads to severe restrictions if thermodynamical consistency is
imposed. However, consistent solutions do exist, both
exponential and power--law.
These solutions respect the second law of thermodynamics and have
positive effective specific entropy. Of course, the phenomenological
theory lacks a microscopic foundation, and it is necessary to 
assume a priori that long--range interactions or some other
mechanism are present in order to maintain the viability of a
thermo--hydrodynamical description during inflation. If the theory 
presented here were applied to non--inflationary processes that
are far from equilibrium (e.g. in astrophysics), then this
limitation would not apply.

\acknowledgements

Thanks to David Jou for useful discussions.
RM thanks the Department of Physics at Barcelona (especially Diego
Pav\'on) and the Department of Mathematics at Natal (especially Sunil
Maharaj) for warm hospitality. RM was supported by research grants
from Natal and Portsmouth Universities, and VM by the
Spanish Ministry of Education under grant PB94-0718.

\end{document}